# SECRETS OF MAGNETO-HYDRODYNAMIC HELL:
## The Solar Circulation Paradox and the Geodynamo

Bertrand C. Barrois[*]

*Institute for Defense Analyses*
*Revised, July 2009*

Differential rotation is widely supposed to be essential for the dynamo effects that sustain solar and planetary magnetic fields, but dynamo effects tend to oppose the flows that drive them, and it is uncertain what drives differential rotation. The relative sign of the differential rotation and meridional circulation is not consistent with simple convection modified by Coriolis forces. We investigate dynamo mechanisms consistent with the observed solar circulation, and discuss how reactive JxB forces would affect such flows. We formulate scaling rules that relate the magnetic field strength to mean rotation and convective heat transport.

Traditional discussions of stellar and planetary dynamos are obscure and conceptually unsatisfactory. The old language of *alpha* and *omega* effects is vague and begs for better definition. (The alpha effect refers to miscellaneous correlations between fluctuating magnetic fields and fluid flows, which are presumed to have cross-terms that regenerate the main magnetic field. The omega effect refers to Coriolis phenomena that convert poloidal into toroidal flows.)

Kinematic dynamo models, which take a time-independent fluid flow as given, tell only half the story. It is widely believed that the dynamo effect depends on differential rotation and/or meridional circulation, which must also be explained. The signs of these flows are critical to the dynamo, and it seems likely that magnetic as well as Coriolis forces play a role in driving and/or orienting them.

Some fifty years ago, Bullard & Gellman took the productive approach of expanding both the magnetic field and the fluid flow in a set of vector spherical harmonics, and of describing the time-evolution of a kinematic dynamo by coupled differential equations. But the limitations of their truncated basis soon became apparent, and ever since, only the foolhardy have dared to simplify.

---

[*] Address: IDA, 4850 Mark Center Dr, Alexandria VA 22311.  E-mail: BBarrois@ida.org

Although the lavish simulations now running on supercomputers appear to contain all the pertinent physics, they are almost too detailed, and it is a challenge to highlight the relevant patterns and correlations, which is what these notes aim to do.

**The Solar Circulation Paradox**

Steady flows are observed on the surface of the sun. Rotation is faster at the equator (25-day period) than at the poles (35-day period). The meridional circulation at the surface is an order of magnitude slower. It goes from the equator toward the poles, with a speed approaching 40 m/s at mid-latitudes.

One is immediately tempted to relate these flows via the Coriolis force, but one must first decide which drives which. Recall that rotations around the *z*-axis carry X to Y but carry Y to -X. The relationship depends on who leads, and who follows.

The relative sign of the flows is inconsistent with simple convective mechanisms. It is easy to see that convection could drive meridional circulation, which brings hot gas up from the depths. But if the poleward meridional circulation were the driver, the Coriolis force would make the rotation slower at the equator and faster at the poles, contrary to fact. If differential rotation were the driver, it should act as a centrifugal pump, drawing fluid outward at low latitudes, and returning it at higher latitudes, exactly as observed. But this raises a further riddle: "What could possibly drive differential rotation?" We will look for explanations in the balance between hydrodynamic forces and reactive side-effects of the dynamo mechanism.

**Fundamental Equations**

The time evolution of the magnetic field is governed by the following equation, which can be derived by assuming that $\mathbf{J} = \sigma(\mathbf{E} + \mathbf{v} \times \mathbf{B})$, and then eliminating $\mathbf{E}$ from Maxwell's equations. (It is safe to drop the displacement term because slow changes do not radiate significant amounts of energy.)

$$\dot{\mathbf{B}} = \operatorname{curl}(\mathbf{v} \times \mathbf{B}) + \operatorname{curl} \tfrac{1}{\sigma\mu} \operatorname{curl}(\mathbf{B})$$

Interpretation: Magnetic field lines are dragged along ("advected") by the moving fluid, and are compressed wherever the fluid converges, but tend to diffuse and dissipate because of finite conductivity. The diffusion term simplifies if the conductivity ($\sigma$) is constant.

$$\dot{\mathbf{B}} + (\mathbf{v} \cdot \nabla)\mathbf{B} = (\mathbf{B} \cdot \nabla)\mathbf{v} - \mathbf{B}\operatorname{div}(\mathbf{v}) + \tfrac{1}{\sigma\mu} \nabla^2 \mathbf{B}$$

The fluid flow is driven by buoyancy, twisted by Coriolis forces, and restrained by magnetic drag. The buoyancy term is written in terms of *excess* enthalpy ($\delta H$) with expansion coefficient $a \equiv [d(\log V)/dH]_P = \alpha/C_P$, where $H$ and $C_P$ are *per gram*.

$$\dot{\mathbf{v}} + (\mathbf{v}\cdot\nabla)\mathbf{v} + \mathbf{\Omega}\times\mathbf{v} = \tfrac{1}{\rho}\mathbf{J}\times\mathbf{B} - \mathbf{g}a\delta H - \tfrac{1}{\rho}\nabla\delta P$$

In our slow-motion approximation, we may assume that the pressure deviation ($\delta P$) is chosen to enforce the incompressibility constraint div($\rho\mathbf{v}$) = 0.

Finally, the excess enthalpy is fed by heat sources and sinks, with ohmic and hydrodynamic contributions.

$$\dot{\delta H} + (\mathbf{v}\cdot\nabla)\delta H = h + \tfrac{1}{\rho\sigma}J^2 + \mathbf{v}\cdot\mathbf{g}a\delta H + \chi\nabla^2\delta H$$

Entropy-based accounting would allow us to omit the $\delta VdP$ term, but requires knowledge of the ambient temperature profile.

$$\dot{\delta S} + (v\cdot\nabla)\delta S = \frac{1}{T_o + \delta T}\left[h + \tfrac{1}{\rho\sigma}J^2 + \chi\nabla^2\delta H\right]$$

The irreversible diffusion term now generates entropy as well as redistributing it, thanks to the second-order correlation between $\delta T$ and $\delta H$. With a little help from integration by parts, the entropy source is seen to be $\chi\langle\nabla\delta S\cdot\nabla\delta S\rangle/C_p$.

The geophysical heat source (*h*) might be a distributed source such as radioactivity, while the sink might be heat conduction across the core-mantle boundary. (A competing hypothesis asserts that the convection is driven by progressive freezing of the inner core. When pure iron freezes out, streamers of less-dense elements have much the same buoyancy as heated material. Although the heat of fusion released by freezing is small, the work done against buoyancy would be vastly greater. In this case, the source would be localized at the inner boundary, ~~and the sink distributed~~.)

Settling ultimately converts gravitational potential energy into heat, which is transported outward to a sink at the outer boundary.

In either case, the equivalent heat transport probably amounts to 1-10 TW. Given our estimate of the adiabatic temperature gradient at *R*(outer) = 3580 km, ~ 1.2 deg/km, heat conduction alone could transport as much as 5 TW, leaving little to be transported by convection. The settling of denser elements over some five billion years could release the gravitational equivalent of ~2 TW at *R*(inner) = 1220 km, and this alternative mechanism is not prone to serious competition from chemical diffusion, which is much slower than heat diffusion mediated by conduction electrons.

In any coarse-grained analysis, thermal and magnetic diffusion are overwhelmed by turbulent transport, termed hyper-diffusivity, but they are essential in principle to enable dissipation at short length scales. It is safe to ignore viscosity entirely, because it does not even enter the MHD stability analysis.

**Energy Conservation**

Energy takes several forms in the geodynamo -- kinetic, thermal, gravitational, and electromagnetic -- the sum of which must be conserved second by second. Moreover, the energy in each of these categories is bounded over the long run. Energy can slosh back and forth, but there can be no steady transfers between categories.

Magnetic drag draws down the kinetic energy of the fluid at a rate of $\mathbf{J}\cdot\mathbf{v}\times\mathbf{B}$, while energy is transferred (non-locally) to the electromagnetic fields at a rate of $-\mathbf{J}\cdot\mathbf{E}$. Taking the difference, we find that ohmic heating must amount to $\mathbf{J}\cdot(\mathbf{E}+\mathbf{v}\times\mathbf{B}) = J^2/\sigma$.

Since kinetic energy remains bounded, we may conclude that the work done by buoyancy must ultimately balance the work done by magnetic drag. This consideration suggests a scaling rule of the form: $\sigma B^2 \langle v^2 \rangle \sim ag\rho\langle vH \rangle \sim agF$, where $F$ denotes the convective heat transport though the system (i.e., flux = power / area). The magnetic field energy is hundreds or thousands of times greater than the kinetic energy of the fluid.

Such arguments do not reveal where the work is done. Poynting's vector can reshuffle electromagnetic energy, and the convection engine can accept heat at one depth while delivering work at another.

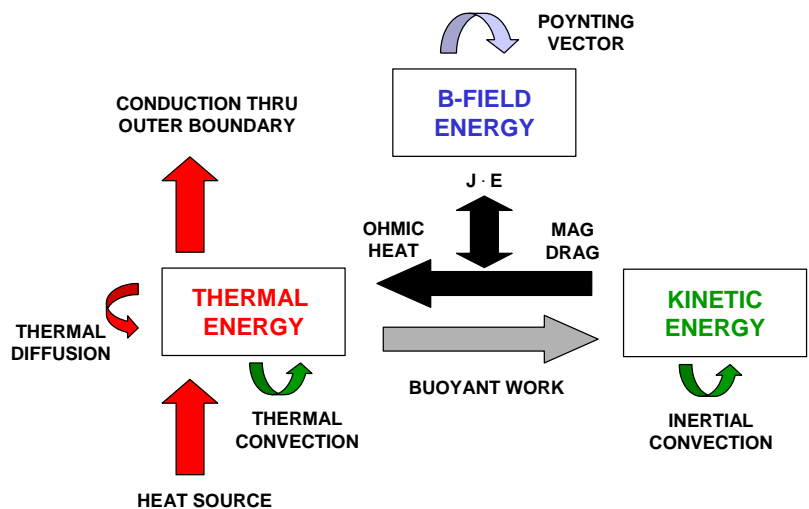

The *dissipation number* represents the ratio of work done by buoyancy over the entire volume (then dissipated by magnetic drag and redeposited) to the net power transported through the system. It may be estimated as $\int dR\, ag \sim 0.65$.

**Magnetic Drag on High-Order Modes**

Magnetic drag is not isotropic. A pure $\sigma(\mathbf{v}\times\mathbf{B})\times\mathbf{B}$ force would always oppose transverse flows, but when the induced currents diverge or dead-end on a boundary, charge separation will occur, and the resulting $\mathbf{E}$ field will offset or cancel $\mathbf{v}\times\mathbf{B}$. We may calculate the drag from a uniform magnetic field on a shear flow with $\mathbf{v} \sim \cos(\mathbf{k}\cdot\mathbf{x})$ as $\mathbf{F} = \vec{\mathbf{P}}\cdot(\mathbf{B}\times(\vec{\mathbf{P}}\cdot(\sigma\mathbf{B}\times\mathbf{v})))$, where the tensor $\vec{\mathbf{P}} = \vec{\mathbf{1}} - \mathbf{k}\mathbf{k}/k^2$ enforces the constraints $\mathrm{div}(\mathbf{v}) = \mathrm{div}(\mathbf{J}) = 0$.

Defining $\beta = \frac{\sigma}{\rho}B^2$, taking 4 gauss for the external dipole field strength extrapolated into the core, and $\leq 0.6$ MS/m for the electrical conductivity, we find $1/\beta$ on the order of a *day*. (This is suspiciously close to the rotation period. Can it be mere coincidence? The sun's ratio of $\beta/\Omega$ is three orders of magnitude times greater.)

The toroidal magnetic field internal to the core has no external manifestations but is thought to be significantly stronger than the external dipole field. The toroidal field would have no effect on parallel toroidal flows such as differential rotation but could be the dominant drag on eddies. We will need the language of vector spherical harmonics to discuss these low-order modes.

**Vector Spherical Harmonics**

Since both the fluid flow ($\rho\mathbf{v}$) and the magnetic field ($\mathbf{B}$) are divergence-free, it is convenient to expand them in vector spherical harmonics. These vector functions come in *toroidal* and *poloidal* series, labeled by the familiar quantum numbers *L*, *M*, and the number of radial nodes. They may be constructed as follows, starting with an arbitrary scalar function *F*:

$$\mathbf{T} = curl(\mathbf{R}F) = -\mathbf{R}\times\nabla F; \qquad \mathbf{S} = curl(\mathbf{T})$$

Note that the toroidal vector has no radial component, but that if $F = f(r)\,Y_{LM}(\theta,\phi)$, the radial component of the poloidal vector will faithfully reproduce the initial scalar: $\mathbf{R}\cdot\mathbf{S} = L(L+1)F$.

The fields and flows obey subtly different boundary conditions. The fluid flow may not cross a boundary, but its parallel components may slip. The magnetic field must join smoothly to a curl-free external field, but its parallel components must taper off unless there is a surface current. We will use the self-explanatory notation "VS(L,M)" (and so forth) to describe patterns consistent with these boundary conditions, where U = steady flow, V = fluctuating flow, B = magnetic field, J = electric current, S = poloidal, and T = toroidal. Since we are only interested in real functions, we will write (L,M,S) and (L,M,C) for the sine and cosine functions of azimuth, when the distinction is significant.

The radial dependence is not indicated. There is no good reason to use spherical Bessel functions; polynomials are more convenient; and either can be used to construct an orthonormal basis. It is easy to integrate polynomial functions over the unit ball (the interior of the unit sphere) using the rule:

$$\langle x^{2a} y^{2b} z^{2c} \rangle_{ball} = 3 \frac{(2a-1)!! (2b-1)!! (2c-1)!!}{(2a+2b+2c+3)!!}$$

Spherical harmonics are convenient because of the spherical boundary conditions, but it should be noted that Coriolis forces break the symmetry and that $L$ is a "bad" quantum number in this problem. Only $M$ and parity are "good" quantum numbers. There cannot be any correlations between the amplitudes of field and flow components with different quantum numbers.

It is amusing to note that patterns of the most prominent parities (with respect to $x,y,z$ or just $z$) form closed sub-algebras. For example, it would be theoretically possible to find solutions purely of VS(even,M) + VT(odd,M) + BS(odd,M) + BT(even,M). But nature seems to have other ideas. The slightest imperfections can mix these patterns with others of opposite parity, and there is no hint of an even-odd parity imbalance in the observable geomagnetic field.

The main magnetic field patterns (with a nominal taper) are as follows:

- BS(1,0) = $[-xz, -yz, (-2 + 2x^2 + 2y^2 + z^2)]$ from $F = z(1 - r^2/2)$
- BT(2,0) = $[yz, -xz, 0](1 - r^2)$ from $F = (x^2 + y^2 - 2z^2)(1 - r^2)$

The most prominent flows are meridional circulation and differential rotation:

- US(2,0) = $[x(1 - x^2 - y^2 - 3z^2), y(same), z(-2 + 4x^2 + 4y^2 + z^2)]$
- UT(odd,0) = $[y, -x, 0]$ (untapered)

The conservation of angular momentum demands that all three VT(1,M) flows remain constant. In a co-rotating frame, they are simply zero.

**Magnetic Drag on Low-Order Modes**

We are now in a position to calculate the magnetic drag on low-order modes. The drag coefficient may be defined as $\dfrac{\langle (\mathbf{v} \times \mathbf{B} + \mathbf{E})^2 \rangle}{\langle v^2 \rangle \langle B^2 \rangle}$. The effects of the pure BT(2,0) field on VS(L,M) and VT(L,M) flows derived from $F = (1 - r^2)\, r^L Y_{LM}$ are shown in the table. It is apparent that magnetic drag is strongest at mid-latitudes, on modes with $M \sim L/2$.

**Magnetic Drag Coefficients due to BT(2,0)**

| VS(L,M) | M=0 | M=1 | M=2 | M=3 | M=4 |
|---|---|---|---|---|---|
| L=1 | 0.123 | 0.106 | | | |
| L=2 | 0.051 | 0.432 | 0.202 | | |
| L=3 | 0.065 | 0.247 | 0.547 | 0.229 | |
| L=4 | 0.072 | 0.177 | 0.415 | 0.557 | 0.222 |
| VT(L,M) | M=0 | M=1 | M=2 | M=3 | M=4 |
| L=1 | 0 | 0.434 | | | |
| L=2 | 0 | 0.271 | 0.510 | | |
| L=3 | 0 | 0.176 | 0.521 | 0.478 | |
| L=4 | 0 | 0.117 | 0.394 | 0.615 | 0.418 |

The calculation says that the toroidal field exerts very little drag on steady meridional circulation. Moreover, VS(L,0) flows will ultimately distort the toroidal field so as to further reduce the drag. In the absence of diffusion and/or turbulent transport, the steady state would have $curl(\mathbf{v} \times \mathbf{B}) = 0$, and $\mathbf{E} = -\nabla \Phi$ would cancel $\mathbf{v} \times \mathbf{B}$. This leaves the external (poloidal) field extrapolated into the core as the dominant drag on steady flows.

**Convective Stability Analysis**

The Bénard analysis of convective instabilities can be heuristically adapted to reveal the modes that are active after the onset of convection. An eigenvector analysis of temperature variations, poloidal, and toroidal flows, coupled by buoyancy and Coriolis forces, suggests that there are three families of modes. In the case of eddies with M>0, there is a marginally unstable convection mode (restrained only by turbulent transport phenomena) as well as two strongly damped flow modes. The marginally stable mode is a "magnetostrophic flow" or "thermal wind", of the sort used by Graeme Sarson to drive kinematic dynamos.

Let $\Theta$ denote the deviation from ambient temperature, $\theta$ the mean super-adiabatic temperature (or composition) gradient, and $\alpha$ the thermal expansion (or buoyancy) coefficient. The linearized equations of motion are

$$\dot{\Theta} = \mathbf{\theta} \cdot \mathbf{v}; \quad \dot{\mathbf{v}} = \alpha \Theta \mathbf{g} - \nabla P + \mathbf{v} \times \mathbf{\Omega} - \vec{\mathbf{\beta}} \cdot \mathbf{v}; \quad div(\mathbf{v}) = 0$$

If we were to neglect the anisotropic effects, namely magnetic drag and Coriolis forces, the unstable modes (with growth rate $\lambda$) would emerge as pure *L*-states from a generalized eigenvalue problem:

$$-\lambda^2 \nabla^2 F = \alpha (\mathbf{\theta} \cdot \mathbf{g}) \frac{L(L+1)}{r^2} F \quad \text{with} \quad v_S = curl(curl(\mathbf{R}F))$$

Both magnetic drag and advection by differential rotation tend to mix (L,M) modes with (L±2,M), provided that M≠0.

If we single-out a specific VS(L,M) mode and neglect its couplings to other poloidal modes of like parity, the truncated eigenvalue problem takes the form shown below, with inhomogeneous driving terms added. Let *S* denote the VS(L,M) velocity, and *T* the VT(L±1,M) velocity. Furthermore, let $\beta_{S,T}$ denote the rate(s) of damping due to magnetic drag, $\Omega$ the effective Coriolis parameter related to the rotation rate, $g\alpha$ the effective buoyancy coefficient, and $\gamma$ the damping due to turbulent heat transport. All parameters are defined to be positive, while *S* & *T* are positive in the observed directions (i.e., poleward at the surface, prograde at the equator.)

$$\frac{d}{dt}\begin{bmatrix} \Theta \\ S \\ T \end{bmatrix} = \begin{bmatrix} -\gamma & \theta & 0 \\ g\alpha & -\beta_S & +\Omega \\ 0 & -\Omega & -\beta_T \end{bmatrix} \begin{bmatrix} \Theta \\ S \\ T \end{bmatrix} + \begin{bmatrix} F_\Theta \\ F_S \\ F_T \end{bmatrix}$$

The effective buoyancy coefficient and Coriolis parameter depend on (L,M) via

$$(g\alpha\theta)_{eff} = \frac{\langle \alpha(\mathbf{v}\cdot\mathbf{\theta})(\mathbf{g}\cdot\mathbf{v})\rangle}{\langle v_S^2\rangle}; \quad \Omega_{eff}^2 = \frac{\langle \mathbf{v}_S \cdot \mathbf{\Omega} \times \mathbf{v}_T\rangle^2}{\langle v_S^2\rangle\langle v_T^2\rangle}.$$

In a thin shell, representative of the sun's convective zone, the value of $(g\alpha\theta)$ is greater for L=2 than for L=1. This seems to hold true in the earth as well, although the temperature and/or composition gradients are poorly known.

The boundedness of the system forbids positive eigenvalues, because they would indicate runaway growth. Hence, turbulence must build to the point where turbulent transport phenomena that mimic diffusion can restrain the most unstable mode, presumably VS(2,0).

$$\gamma > \frac{g\alpha\theta\,\beta_T}{\beta_S\beta_T + \Omega^2}$$

(A similar analysis applies to high-order modes characterized by wave-vector **k**. Careful calculation shows that at every latitude $\psi$, there exists one direction of **k** at which magnetic drag is ineffectual and the undamped growth rate approaches $\sin(\psi)\sqrt{g\alpha\theta}$, but this exceptional case does not seem to set any time scales.)

The concept of eddy viscosity is somewhat bogus, because genuine diffusion coefficients have units of $L^2/T$, whereas Kolmogorov's theory of turbulence maintains that power spectra and correlation times can be derived by dimensional analysis from a parameter with units of $L^2/T^3$, which represents the power per unit mass cascading from low-order to high-order eddy modes. It follows from dimensional analysis that eddy decay rates scale as the 2/3 power, not the square, of spatial frequency.

If we swallow our misgivings about the effective diffusion coefficient $\chi$, we can derive an approximate relationship between $\gamma$ and $v$. Taking $\gamma \sim \chi(\pi/\Delta R)^2$ and $\theta \sim F/\rho\chi C_P$ midway between the inner and outer boundaries, we find $\chi \sim v(\Delta R/\pi)$. The damping time constant $1/\gamma$ must be on the order of centuries, whereas $1/\beta$ is on the order of hours to days, depending on the strength of the toroidal magnetic field.

These arguments allow us to estimate the Nusselt number, which represents the renormalization of thermal conductivity by turbulent transport. According to accepted heuristics, the local temperature gradient should diverge as $\theta(x) \sim 1/x$ when approaching a boundary, but the divergence cuts off where conduction takes over from convection. The average temperature gradient $\Delta T/\Delta R$ will exceed the central gradient $dT/dR$ by a logarithmic factor of order $\sim \log(\chi/\chi_o)$. (Beware unknown factors of order unity.)

The inverse matrix reveals the flows in response to various kinds of zero-frequency driving terms. (Note that $det < 0$, and that a positive entry signifies positive response. The relationship of poloidal to toroidal responses is considered "normal" if $S/T \sim -\beta_T/\Omega < 0$, which holds for thermally and/or poloidally driven eddies.) By this definition, the paradoxical solar circulation is considered "abnormal".

$$\begin{bmatrix} \Theta \\ S \\ T \end{bmatrix} = \frac{-1}{\det} \begin{bmatrix} (\beta_S\beta_T + \Omega^2) & \theta\beta_T & \theta\Omega \\ g\alpha\beta_T & \gamma\beta_T & \gamma\Omega \\ -g\alpha\Omega & -\gamma\Omega & (\gamma\beta_S - g\alpha\theta) \end{bmatrix} \begin{bmatrix} F_\Theta \\ F_S \\ F_T \end{bmatrix}$$

The sign of a toroidal response to a toroidal driver (the lower-right entry) is ambiguous. If $\gamma$ barely satisfies the stability bound, the net coefficient could be negative, and a prograde force could produce a "perversely" retrograde response, or vice versa. (Think of a sailboat tacking against the wind.)

"Perversity" notwithstanding, the ratio of meridional circulation to differential rotation would be strictly "normal" in the limit of marginal stability, contrary to nature. (We gather that the system must operate close, but not too close, to the stability bound.)

The "perverse" response diverges as $\gamma$ approaches the stability bound and the determinant approaches zero. What limits it? We might argue that runaway differential rotation over-amplifies the BT(2,0) field, thereby self-quenching. (Once again, the system must operate close, but not too close, to the stability bound.)

We may try to reproduce the observed relationship with a combination of drivers. It should be self-evident that thermal drivers based on $J^2/\sigma C_P$ can be neglected given high conductivity. The relative importance of different drivers involves the dissipation ratio, the magnetic diffusion time scale, and the magnetic drag parameter:

$$\frac{F_\Theta}{F_S} \frac{S}{\Theta} \sim \left(\frac{J^2}{JB}\right)\left(\frac{\alpha g}{\sigma C_P \beta_S}\right) \sim \frac{agR}{\sigma \mu R^2} \frac{1}{\beta_S} \ll 1$$

We now have a 2x2 matrix. The observed combination of slow meridional circulation and fast differential rotation demands that $F_S/F_T \sim -\Omega/\beta_T$, and there is nothing "perverse" about the sign of the required toroidal driver.

$$\begin{bmatrix} F_S \\ F_T \end{bmatrix} = \begin{bmatrix} (\beta_S - g\alpha\theta/\gamma) & -\Omega \\ +\Omega & \beta_T \end{bmatrix} \begin{bmatrix} S \\ T \end{bmatrix}$$

**Effects of Steady Flows**

It is often said that conducting fluids drag field lines, but this statement is naive. If the core rotated as a rigid body, it would have no such effect on a uniform field parallel to the axis of rotation, because charge separation would establish an **E** field that cancelled **V**×**B**. This follows from the fact that (**Ω**×**R**)×**B** is curl-free and can be canceled by $\nabla\Phi$.

This would also be true of rotation-on-cylinders, but the effect of latitude-dependent differential rotation or rotation-on-planes is to shear the field lines into hairpins, thereby creating a strong toroidal field, predominantly BT(2,0).

It is possible to set up an idealized problem by abandoning spherical geometry in favor of straight lines of flow. The effect of a *z*-directed flow on an *x,y*-directed field, both z-independent, is to create an additional z-directed field that may be calculated from $\chi \nabla^2 B_z = -(\mathbf{B} \cdot \nabla) v_z$.

The effect of meridional circulation is to distort the field lines with a touch of BS(3,0) and ultimately to roll them up. It is possible to solve an idealized "jelly roll" problem exactly, again by abandoning spherical geometry, with Hankel functions of complex argument, but the solution is physically unrealistic because the field strength is amplified by a factor of $\exp(2\pi) = 535.5$ per turn of the spiral, thereby imposing prohibitive magnetic drag. (This factor is suspiciously close to the number of voting members of the U.S. Congress. Can it be mere coincidence?) One valuable lesson from the jelly roll model is that the circulation time cannot be much shorter than the diffusion time. The exact time-independent solution takes the form:

$$\text{Circulation:} \quad \mathbf{v} = (\omega y, -\omega x, 0)$$
$$\text{External:} \quad B_x = 1; \quad A_z = y = r \sin(\varphi)$$
$$\text{Internal:} \quad \mathbf{B} = \text{curl}(\mathbf{A}); \quad \chi \nabla^2 A_z = i\omega_z A_z$$
$$A_z = \text{Re}[\ldots H_1(r\sqrt{i\omega/\chi}) \exp(i\varphi)]$$

(The coefficient must be chosen to join smoothly to the external field. A central singularity is inevitable but inoffensive because the total field energy is log-divergent.)

**Effects of Eddies**

It is possible to make up hypothetical VS(L,M) eddies accompanied by Coriolis-twisted VT(L±1,M) eddies and to calculate their combined effects on the magnetic field. These eddies may be somewhat unrealistic in terms of latitude and radial profiles, because $L$ is a "bad" quantum number. (Strictly speaking, we should be considering uncorrelated normal modes, as defined by the Karhunen-Loeve decomposition, involving diagonalization of a covariance matrix of temperatures and/or velocities at many points. It is difficult to identify these normal modes *a priori*, except in translationally invariant systems.)

A key step in constructing the Coriolis-twisted sidekick is to extract the *solenoidal* part of $\mathbf{v} \times \mathbf{\Omega}$. This involves subtracting the gradient of a scalar, chosen to cancel any divergence and/or flow through the boundary. Computationally, the best method is to construct a complete, orthonormal basis of solenoidal functions of sufficient degree. Unless the original pattern is a mixture of many different harmonics, the functions can be generated by repeated application of the operator $(curl)(1-r^2)(curl)$, followed by Gram-Schmidt orthogonalization. Double-precision arithmetic is essential.

The geometric effect of a VS(L,M,C) eddy is to distort BS(1,0) to a linear combination of BT(L,M,S) and BS(L±1,M,C). The same eddy can then act a second time to regenerate BS(1,0). The same is true for the accompanying VT(L±1,M,C) eddy, which distorts BS(1,0) to a combination of BT(L,M,C) et cetera. However, it turns out that the sign of the regeneration coefficient is negative. Instead of amplifying the field, two-stage mechanisms act to diffuse it.

Only VS×BT and VT×BS terms contribute to the dipole moment. Since VT×BT is purely radial, it cannot generate a JT(1,0) current loop. VS×BS is also impotent, for a subtler reason. If $\sigma/\rho$ were uniform, the contribution of VS×BS to the dipole moment would integrate to zero. (Quick proof: Consider the M=0 case, in which all flow and field lines form closed loops in a plane. If a flow line crosses a field line outbound, it must cross the same field line inbound elsewhere. The result can be extended to M>0 using Wigner's theorem. Since the Clebsch-Gordan coefficients for (L,M; L±1,-M; 1,0) do not vanish when M=0, we may conclude that the universal coefficient does.)

Even if $\sigma/\rho$ were significantly non-uniform, all reasonable VS×BS mechanisms (in which the conductivity is a function of depth and/or temperature) seem to produce damped oscillations. The picture below shows the poloidal flow carrying nested field lines around in a grand tour. There is no current density at the center of the nested field loops to counter ordinary ohmic decay. Moreover, the prograde current cannot cancel the retrograde current due to subtle variations of conductivity because they are at the same depth.

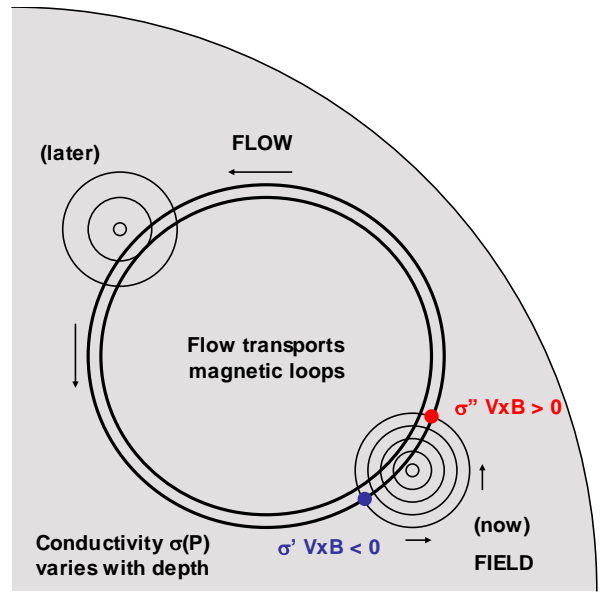

**Pictorial Proof of Cowling's Theorem**

These results are consistent with Cowling's Theorem, which maintains with great generality that axisymmetric flows cannot sustain axisymmetric fields, even in the time-dependent case. The CG coefficient for (L,M; L,-M; 1,0) is proportional to *M* and vanishes for M=0. Second-order coefficients will be seen to scale precisely as $M^2$.

We may define dipole "degeneration" coefficients for **B** = BS(1,0) as follows:

$$\frac{\langle [-y, x, 0] \cdot (\mathbf{v} \times curl(\mathbf{v} \times \mathbf{B})) \rangle}{\langle v^2 \rangle \langle B^2 \rangle^{1/2}}$$

**Dipole Dissipation Coefficients, Two-Stage**

| VS(L,M) | M=0 | M=1 | M=2 | M=3 | M=4 |
|---|---|---|---|---|---|
| L=1 | 0 | -2.08 | | | |
| L=2 | 0 | -0.75 | -3.01 | | |
| L=3 | 0 | -0.39 | -1.59 | -3.57 | |
| L=4 | 0 | -0.25 | -1.54 | -2.23 | -3.96 |

**Three-Stage Dynamo Mechanisms**

Although two-stage mechanisms seem utterly hopeless, some three-stage mechanisms can produce positive regeneration. For example, an "alpha-omega" effect:

- Steady differential rotation pre-distorts BS(1,0) to BT(2,0).

- A VS(L,M) eddy distorts BT(2,0) to BS(L±1,M)

- Its VT(L±1,M) sidekicks distort BS(L±1,M) to regenerate BS(1,0).

- Intermediate BT(L,M) patterns would be allowed by parity, but direct calculation shows that they are not generated. Moreover, VS×BT terms cannot contribute to the dipole moment.

The eddies can act in either order with identical effect. The catch is that some high-M eddies produce strongly negative regeneration, and we would need to explain them away. As we have seen, magnetic damping is actually stronger on medium-M eddies at mid-latitudes, but the ratio of nonlinear drivers is unknown.

The figure illustrates the current component that generates the dipole moment, averaged with respect to longitude. (Prograde is regenerative; retrograde degenerative.) The predominantly prograde current loops are characteristically split between the hemispheres, but the sign of the octupole (L=3) moment is not immediately obvious.

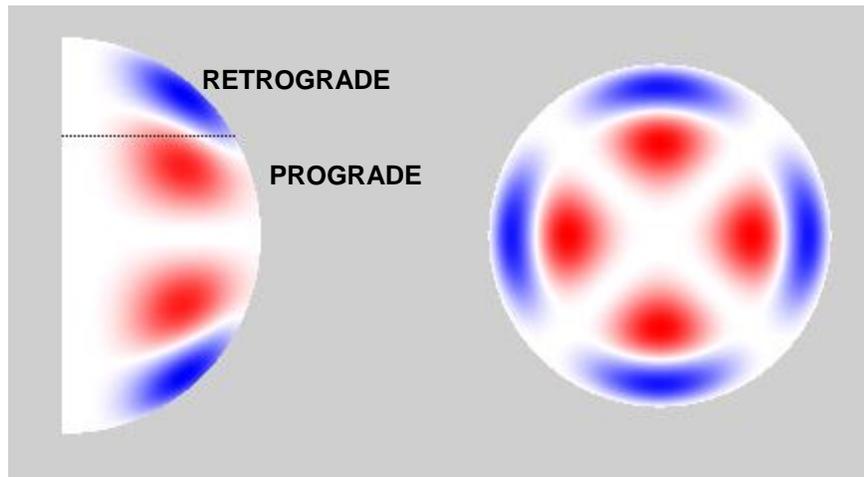

**Azimuth-Averaged Toroidal Currents for L=3, M=2, in Vertical Plane,
and Actual Azimuthal Distribution in a Mid-latitude Horizontal Plane**

We may define dipole regeneration coefficients for BS(1,0) as follows, where $\mathbf{u}$ = UT(3,0), $\mathbf{v}$ = VS(L,M), and $\mathbf{v'}$ = *solenoidal* ($\mathbf{v}\times\mathbf{\Omega}$):

$$\frac{\langle [-y, x, 0] \cdot (\mathbf{v}' \times curl(\mathbf{v} \times curl(\mathbf{u} \times \mathbf{B}))) \rangle}{\langle v^2 \rangle \langle u^2 \rangle^{1/2} \langle B^2 \rangle^{1/2}}$$

**Dipole Regeneration Coefficients, Three-Stage, UT-VS-VT**

| UT-VS-VT | M=0 | M=1 | M=2 | M=3 | M=4 | M=5 | M=6 | M=7 | M=8 |
|---|---|---|---|---|---|---|---|---|---|
| L=1 | 0 | -0.223 | | | | | | | |
| L=2 | 0 | 0.128 | -0.726 | | | | | | |
| L=3 | 0 | 0.120 | 0.152 | -1.187 | | | | | |
| L=4 | 0 | 0.098 | 0.263 | 0.089 | -1.152 | | | | |
| L=5 | 0 | | | | | | | | |
| L=6 | 0 | 0.070 | 0.227 | 0.386 | 0.381 | -0.074 | -1.849 | | |
| L=7 | 0 | | | | | | | | |
| L=8 | 0 | 0.042 | 0.168 | 0.333 | 0.483 | 0.542 | 0.391 | -0.171 | -1.868 |

A competing "alpha" effect (with opposite sign) that does not depend upon differential rotation could be responsible if high-M eddies are dominant:

- Steady meridional circulation pre-distorts BS(1,0) to BS(3,0).
- A VS(L,M) eddy distorts BS(3,0) to BT(L,M).
- The same eddy distorts BT(L,M) to BS(1,0).

**Dipole Regeneration Coefficients, Three-Stage, US-VS-VS**

| US-VS-VS | M=0 | M=1 | M=2 | M=3 | M=4 | M=5 | M=6 | M=7 | M=8 |
|---|---|---|---|---|---|---|---|---|---|
| L=1 | 0 | 1.521 | | | | | | | |
| L=2 | 0 | -0.306 | 2.451 | | | | | | |
| L=3 | 0 | -0.244 | -0.200 | 2.467 | | | | | |
| L=4 | 0 | -0.184 | -0.486 | -0.158 | 2.047 | | | | |
| L=5 | 0 | | | | | 1.448 | | | |
| L=6 | 0 | -0.116 | -0.418 | -0.764 | -0.915 | -0.539 | 0.790 | | |
| L=7 | 0 | | | | | | | 0.133 | |
| L=8 | 0 | -0.080 | -0.314 | -0.648 | -1.023 | -1.337 | -1.464 | -1.232 | -0.502 |

The high-M current loops are concentrated at low latitudes, and the sign of the octupole (L=3) moment seems consistent with the observed multipole expansion of the geomagnetic field, despite the retrograde loop near the boundary.

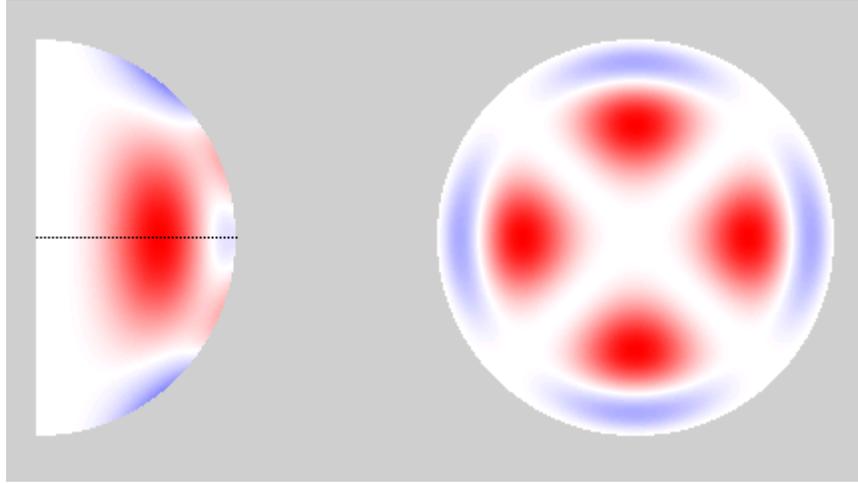

**Azimuth-average Toroidal Currents for L=2, M=2, in Vertical Plane,
and Actual Distribution in Equatorial Plane. Alpha mechanism.**

By themselves, the coefficients give an incomplete picture of relative strength, because the ratio of UT×VS×VT to US×VS×VS effects is enhanced by the ratio of steady UT(2,0) to US(1,0) flows, which is a mystery, but also suppressed by the ratio of VT(L±1,M) to VS(L,M) eddies, roughly $\Omega/\beta_T$. (Further arguments will suggest that the mystery ratio is just big enough to make the two dynamo mechanisms competitive.)

For reasons that will soon become perfectly clear, the steady flow must act either *before* or *after* the rapidly fluctuating flows, to avoid suppression of high-frequency effects. But steady differential rotation cannot act *last*, because UT(odd,0)×BS(even,0) has no azimuthal component to regenerate JT(1,0).

**Time Dependence**

The effect of a fluctuating eddy is to produce a fluctuating magnetic field. We may single out a Fourier component: $v(t) \sim v_o \cos(\omega t)$. Suppose that the composite eddy distorts a steady BT(2,0) to BT(L,M) to BS(1,0). Let *A,B,C* denote the amplitudes of these field modes. Taking the ordinary decay rates of these modes into account, we have coupled evolution equations:

$$\dot{B} = \zeta v(t) A - \lambda B; \quad \dot{C} = \xi v(t) B - \lambda' C$$

Then taking $A(t) = 1$, we find $B(t) = \zeta v_o \dfrac{\omega \sin(\omega t) + \lambda \cos(\omega t)}{\omega^2 + \lambda^2}$. The in-phase term combines with $v(t)$ to regenerate a steady field: $\langle C \rangle = \dfrac{\zeta \xi v_o^2}{\lambda'} \dfrac{\lambda}{\omega^2 + \lambda^2}$. Since the fluctuating flow is not truly periodic, we should average over its bandwidth $\Gamma$, roughly the reciprocal of its correlation time.

$$\left\langle \frac{\lambda}{\omega^2 + \lambda^2} \right\rangle = \int d\omega \frac{\Gamma/\pi}{\omega^2 + \Gamma^2} \frac{\lambda}{\omega^2 + \lambda^2} = \frac{1}{\Gamma + \lambda}$$

It seems ironic that the decay rate $\lambda$ plays a key role in regeneration but ultimately drops out of the numerator. Without it, $B(t)$ and $V(t)$ would be strictly out of phase.

In summing over active eddy modes, we should get something commensurate with the decay rate due to pseudo-diffusion: $o(\gamma) \sim \sum \zeta \xi v^2 / \Gamma$. However, this will be suppressed by a factor of $\Omega/\beta_T$ if it depends on cross-terms between VS and VT.

**A Marvelous Toy**

It is amusing and instructive to consider what would happen in a two-stage model if the regeneration coefficient were positive. Conflating $A$ with $C$, we get

$$\dot{B} = \zeta v(t) C - \lambda B; \quad \dot{C} = \xi v(t) B - \lambda' C$$

Were it not for the decay terms, $v(t)$ would simply drive any (B,C) state along a hyperbola of constant (say positive) $B^2/\zeta - C^2/\xi$. Assume that the state were driven far out along the asymptote. Given $\lambda' > \lambda$, the decay terms could then pull the state to a new hyperbola, so that the B-intercept would grow. Paradoxically, one drives this system by damping it, and it naturally settles into the Methuselah mode.

Unfortunately, this is a poor representation of the three-stage process, because the stage that converts BS(1,0) into BT(2,0) is slow, and the latter is virtually constant.

**Reactive Force Terms**

The steady meridional circulation and differential rotation will be affected by the dynamo's own **J**×**B** reactive force terms.

Suppose that a VS(L,M) eddy distorts BT(2,0) to **B'**, whereas its toroidal sidekick VT(L±1,M) distorts BT(2,0) to **B''**. The $F_S$ driver for meridional circulation comes primarily from **J'**×**B'**+**J''**×**B''** whereas the $F_T$ driver for differential rotation comes from **J'**×**B''**+**J''**×**B'**.

-17-

The driver coefficients have been normalized to equal values of $\langle v_S^2 \rangle$ and $\langle u^2 \rangle$, where **u** may be either US(2,0) or UT(3,0). Their signs are completely consistent.

**Drivers for Steady UT(3,0) and US(2,0) Circulation**

| Diff. Rot. | M=0 | M=1 | M=2 | M=3 | M=4 |
|---|---|---|---|---|---|
| L=1 | 0 | -6.0 | | | |
| L=2 | 0 | -16.1 | -13.9 | | |
| L=3 | 0 | -14.4 | -45.9 | -17.5 | |
| L=4 | 0 | -6.7 | -63.5 | -66.5 | -18.0 |
| Mer. Circ. | M=0 | M=1 | M=2 | M=3 | M=4 |
| L=1 | 4.7 | 7.2 | | | |
| L=2 | 25.3 | 9.9 | 1.2 | | |
| L=3 | 23.2 | 24.2 | 10.5 | 1.4 | |
| L=4 | 19.2 | 27.2 | 21.4 | 10.5 | 6.3 |

Direct calculation confirms that the toroidal force terms oppose the observed differential rotation, which is obviously not what we want. And the poloidal force terms drive meridional circulation in the observed direction, which is paradoxically not what we want, because we need something to slow it down.

Referring back to the 2x2 matrix relationship between drivers and responses, we find that the observed combination of slow meridional circulation and fast differential rotation demands that $F_S / F_T \sim -\Omega / \beta_T$. (And we now pull out what is left of our hair.)

The conservation of energy illuminates the paradox. One of our candidate dynamo mechanisms involves a conspiracy of differential rotation, poloidal, and toroidal eddies. Conservation of energy requires that the dynamo mechanism oppose one or more of the motions that drive it, but perhaps not all three. Indeed, direct calculation of the toroidal drivers confirms that they *oppose* the observed differential rotation.

As for magnitudes, the matrix relationship between drivers and responses demanded that $|F_S / F_T| \sim \Omega / \beta_T > \Omega / \beta_{ext}$, which could potentially be reconciled with $|F_S / F_T| \sim B'/B'' \sim (S/T)_{eddy} \sim \beta_{int} / \Omega$. But that is small consolation when the signs are wrong.

The second dynamo mechanism involves a conspiracy of meridional circulation and poloidal eddies, but BT(2,0) plays no role. These flows distort BS(1,0) to BS(3,0) and thence to **B**, and indeed we find that the reactive **J**×**B** force opposes the meridional circulation. However, the **B**' field derived from distorting BT(2,0) is even stronger, so **J**'×**B**' overwhelms **J**×**B** and drives meridional circulation in the observed direction, which is not what we want.

**What drives differential rotation?**

If not the dynamo's own reactive forces, then what? We must look to hydrodynamic terms: $(\mathbf{v}_S \cdot \nabla)\mathbf{v}_T + (\mathbf{v}_T \cdot \nabla)\mathbf{v}_S$. These advection terms would sum to zero by symmetry if all *M*-states of given *L* had equal kinetic energy, but that seems unlikely. The high-*M* eddy modes help to drive differential rotation in the observed direction, whereas low-*M* modes (M<<L) have the opposite effect.

Eddy velocities should be fastest in the equatorial plane because the main BT(2,0) field goes through zero there.

**Summary: Are we well done yet?**

Do we know enough to predict the magnitude of the externally observable dipole field? Energy balance tells us that $\beta_{int} v^2 = \sigma B_{int}^2 v^2 \sim agF$, but how are we to separate the eddy velocity from the internal field strength, and then relate the latter to the external field strength?

Naive dimensional analysis is ultimately inconclusive because we have a choice among three very different time scales: rotational $\Omega$, thermodynamic $\Xi \equiv \sqrt[3]{agF/\rho R^2}$, and magnetic diffusion $\Delta \equiv 1/\sigma\mu R^2$. The simplest possible scaling rule would be $\beta = (\Omega)^{\cdots}(\Xi)^{\cdots}(\Delta)^{\cdots}$, with exponents that sum to one, but dimensional analysis admits arbitrary functions of dimensionless ratios: $\beta = \Omega\, f(\Xi/\Omega, \Delta/\Omega)$.

We would need a system of interlocking relationships among these three known parameters and four unknown parameters: internal field strength, external field strength, differential rotation, and eddy velocity. But unfortunately, our current understanding of differential rotation and the regeneration mechanism are too sketchy.

**The Bootstrap Paradox**

*In the beginning, God created the heavens and the earth.* But what about Hell? The usual answer is that the dynamo mechanism must have amplified a weak magnetic seed field, but this is easier said than proven.

The analysis of convective instability says that Coriolis forces are a stabilizing influence, and that the growth rate $\gamma \to 0$ as the magnetic drag $\beta \to 0$.

Moreover, Taylor's Theorem says that, absent dissipative or advection forces, Coriolis forces should regiment the flow into *z*-independent columns unfavorable to dynamo action. (Quick proof: Take the *curl* of $0 = \mathbf{v} \times \mathbf{\Omega} - \frac{1}{\rho}\nabla P$.) It takes either turbulence or magnetic drag to buck the tyranny of Taylor columns. If the dynamo was inefficient at start-up, we must explain how it managed to overcome ohmic diffusion.

The scaling rules formulated above suggest that Coriolis dominance does not suppress the magnetic field, and this may point the way to a resolution of the paradox.

**Dominance of Low-Order Eddy Modes**

In every discussion of turbulent eddies, it is obligatory to quote Jonathan Swift:

*Great fleas have little fleas upon their backs to bite 'em,*
*And little fleas have lesser fleas, and so ad infinitum.*
*And the great fleas themselves, in turn, have greater fleas to go on;*
*While these again have greater still, and greater still, and so on.*

We have mentioned that pseudo-diffusion coefficient is a sum over eddy modes, without inquiring about the contribution of individual modes.

The existence of multiple time scales could undermine the elementary argument from dimensional analysis commonly used to justify the Kolmogorov spectrum, but the spectrum nevertheless seems correct because only one family of convective modes is unstable, while the others are strongly restrained. The Kolmogorov spectrum derives from a parameter with units of $L^2/T^3$, here $\gamma v^2$, and the spatial frequency *k*. We want the combination that has units of $L^2/T$:

$$\sum V^2/\Gamma \sim \left\langle \gamma v^2 \right\rangle^{1/3} \int \frac{d^3k}{k^3} k^{-4/3}$$

The integral over spatial frequency has a low-*k* cutoff imposed by the dimensions of the system and converges slowly at high *k*.

Unfortunately, this argument gives no clue about the distribution with respect to angular quantum numbers. The activity in any mode can be estimated from the quotient of a nonlinear driving term, divided by the net decay rate or effective bandwidth of that mode. For high-order modes, the decay rate ($\gamma$) due to pseudo-dissipation scales as $k^1$, and overwhelms the growth rate ($g\alpha\theta/\beta$). The fact that magnetic drag is very low for axisymmetric (M=0) flows and highest at mid-latitudes (M~L/2) becomes irrelevant.

We dare not conclude that isotropy is restored for high-order modes, however, because **v** still has preferred directions on account of buoyancy and couples to **k** via $(\mathbf{v}\cdot\nabla)\Theta$.

**Field Reversals**

No discussion of dynamo mechanisms is complete without a bit of irresponsible speculation on the subject of field reversals.

We have seen that eddy modes participate in the candidate dynamo mechanisms with inconsistent signs. If there exist eddy modes that contribute positively to pseudo-diffusion but negatively to regeneration, they may be characterized as "anti-dynamo" modes that might be responsible for flipping the dipole field.

*Hypothesis*: An *n*-sigma fluctuation of anti-dynamo modes can provoke a field reversal. The frequency of occurrence is $\exp(-n^2/2)$ per correlation interval $\tau \sim 1/\Gamma$.

Since an anti-dynamo mode must act via a two-stage process to cause a reversal, the effect of a fluctuation is proportional to $n^2v^2\tau$. It follows that an *n*/2-sigma fluctuation sustained for $4\tau$ would be equally potent, but it would occur just as rarely. Several anti-dynamo modes might cooperate to provoke a reversal, but the basic result is the same, and $v^2\tau$ should be understood as a sum over all anti-dynamo modes.

How are we to explain why the sun's dipole field has been oscillating in recent centuries on a fairly regular 22-year cycle, whereas the earth's field reverses at erratic intervals? (There were no reversals during the cretaceous superchron, which lasted almost 40 million years. The average interval between reversals then slid gradually to 200,000 years, but there have been no reversals during the last 780,000 years, known as the Brunhes epoch. As for the sun, there were no spots during the Maunder minimum, 1645-1715 AD.)

One hypothesis is that the sun's field configuration "rotates" between two states under the influence of a persistent anti-dynamo flow mode. The trouble is that all known flow modes produce externally observable BS(L,M) fields, along with confined BT(L,M) fields, but other BS(L,M) modes are not especially prominent during reversals. For example, VS(1,1) could interconvert BS(1,0) and BT(1,1), but it would also produce measurable BS(2,1). It has been reported that the BS(1,0) field is appreciably contaminated with BS(2,2), which indicates the influence of VS(3,2) and VT(2,2).

A more elaborate hypothesis is that some coherent combination of BS(L±even,M) and BT(L±odd,M) could "rotate" the main field to a purely toroidal combination of BT(L±even,M) patterns and back. The toroidal fields would be strictly internal.

Although axisymmetric (M=0) flows are sometimes capable of inducing damped oscillations of the dipole moment, their influence in the sun can be ruled out on several grounds. First, $\sigma/\rho$ is essentially constant, except for a logarithmic factor, and second, the meridional circulation time is an order of magnitude shorter than the 22-year oscillation period. (Consolation: There is evidence of a weak 2-year cycle.)

**Summary: What needs to be done?**

The mechanisms that we have proposed are not robust. There is a tug-of-war between opposing US-VS-VS and UT-VS-VT effects, which seem to be of comparable strength, the winner to be decided by the relative amplitude of various eddy modes, and the balance between regeneration and turbulent diffusion is equally delicate.

The picture painted here raises a multitude of hard issues of nonlinear dynamics that I will leave for you to ponder:

- Symmetry arguments alone cannot identify the normal modes of the system. Can they be identified by approximations short of detailed simulation?
- Can the power spectrum and time scales of the dominant low-order flow modes be estimated by studying a severely truncated nonlinear system?
- How close to the stability bound does the dynamo operate?
- Dimensional analysis is a blunt instrument. Can the saturation values of the internal and external Elsasser ratios be pinned down?
- How do the mean values of steady fields compare to the RMS values of fluctuating fields? Is it legitimate to calculate magnetic drag on the assumption that steady BT(2,0) is dominant?

# APPENDIX: THERMODYNAMICS OF HELL

Table ... summarizes standard estimates of geophysical parameters in the liquid outer core. These are derived by integrating the equations of hydrostatic equilibrium and gravity, and by estimating the equation of state $P(\rho)$. The measured speed of sound reveals the adiabatic derivative, $u^2 = (dP/d\rho)_S$, but the density has two discontinuities, and the actual temperature gradient is slightly super-adiabatic.

The temperature is crudely bounded by data on phase transitions. (The Clausius-Clapeyron slope, $dT/dP = \Delta V/\Delta S$, is complicated by second derivatives of $\Delta G(P,T)$, namely the differences of specific heats, thermal expansions, and compressibilities.)

**Conditions in the Outer Core**

| Parameter | Symbol | Outer Boundary | Inner Boundary | Units |
|---|---|---|---|---|
| Radius | r | 3480 | 1220 | km |
| Gravity | g | 10.7 | 4.4 | m/s$^2$ |
| Density | $\rho$ | 9.9 | 12.6 | g/cm$^3$ |
| Pressure | P | 135 | 330 | GPa |
| Speed of Sound | u | 8.05 | 10.25 | km/s |
| Temperature | T | > 2500 | < 7500 | K |
| Adiabatic gradient | $\nabla T$ | 1.2 (est) | 0.9 (est) | deg/km |

Table ... summarizes some low-pressure thermodynamic data important to MHD, at room temperature for the solids, but just above the melting point for liquid iron. Many significant parameters of molten iron have never been measured, and the corresponding values for solid iron are quoted as crude surrogates.

**Low-Pressure Thermodynamic Data**

| Parameter | Symbol | Solid Fe | Liquid Fe | Solid Hg | Liquid Hg | Units |
|---|---|---|---|---|---|---|
| Temperature | -- | room | 1811 (MP) | 234 (MP) | room | Kelvin |
| Molar Mass | m | 55.85 | 55.85 | 200.6 | 200.6 | g/mole |
| Molar Volume | V | 7.1 | 8.0 | 14.13 | 14.80 | cc/mole |
| Density | $\rho$ | 7.9 | 7.0 | 14.20 | 13.55 | g/cm$^3$ |
| Bulk Modulus | -VdP/dV | 171 | ??? | 24.75 | 27.00 | GPa |
| Speed of Sound | u | 5.15-5.95 | ??? | ??? | 1.41 | km/s |
| Heat Capacity | $C_P$ | 6.0 | 9 (solid) | 6.60 | 6.65 | cal/deg/mole |
| Entropy of Fusion | $\Delta S$ | 2.8 | 2.8 | 2.33 | 2.33 | cal/deg/mole |
| Thermal Expansion | $\alpha$ | 35 | 80 | 48 | 180 | 10$^{-6}$ deg$^{-1}$ |
| Thermal Conductivity | -- | 80 | 35 (solid) | ??? | 8.34 | W/m/deg |
| Electrical Conductivity | $\sigma$ | 10.0 | 0.6 (est.) | ??? | 1.04 | MS/m |
| Grüneisen Parameter | VdP/TdS | 1.7 | 3.0 (est.) | 0.6 | 2.6 | -- |

The heat capacity of an ideal harmonic solid should be about 6 cal/deg/mole. Solid iron has precisely this value at room temperature, but its heat capacity continues to rise, peaking near the Curie temperature at 1044 K and approaching 9 cal/deg/mole for γ-iron at 1500 K. Much of the excess heat capacity is associated with fluctuations of the magnetic order parameter.

The speed of sound in solid iron depends on how it is measured. The speed of sound in a thin fiber depends on Young's modulus (~210 GPa) but the speed of sound in bulk matter depends on the $C_{11}$ modulus (~280 GPa) rather than the bulk modulus for isotropic compression (~170 GPa). Young's modulus presumes no transverse stress, whereas $C_{11}$ presumes no transverse strain. These moduli can be related by Poisson's ratio, $p \sim 0.29$, defined as the ratio of transverse expansion to longitudinal compression under unidirectional stress.

$$\frac{Young's}{Bulk} = 3(1-2p) \qquad \frac{C_{11}}{Young's} = \frac{1-p}{1-p-2p^2}$$

Another key parameter is the thermal expansion coefficient, but it is futile to extrapolate low-pressure data, because thermal expansion at great depths would entail a huge amount of work against pressure. However, this coefficient can be related to better known parameters by the chain rule. (For solids, use the bulk modulus in place of $mu^2$.)

$$\alpha = \left(\frac{dV}{VdT}\right)_P = \frac{1}{V}\left(\frac{TdS}{dT}\right)_P \left(\frac{-dV}{VdP}\right)_S \left(\frac{VdP}{TdS}\right)_V = \frac{C_P}{mu^2}\left(\frac{VdP}{TdS}\right)_V$$

The dimensionless ratio $(VdP/TdS)_V$, known as Grüneisen's parameter, varies widely, but there is reason to believe that it varies slowly with density. Some of the data needed to calculate it for molten iron are missing. Using rough surrogates, we can estimate the mean adiabatic temperature gradient as ~1.1 deg/km, which leaves ample room for a super-adiabatic gradient to drive the convection. (The derivation uses the Maxwell relationship that follows from $dH = TdS + VdP$, as well as the chain rule.)

$$\left(\frac{dT}{dz}\right)_S = g\rho\left(\frac{dT}{dP}\right)_S = g\rho\left(\frac{dV}{dS}\right)_P = \left(\frac{VdP}{TdS}\right)_V \frac{gT}{u^2}$$

The physical basis of Grüneisen's parameter relates to the anharmonicity of the interatomic potential, and the actual value is consistent with stiff short-range repulsions. Consider a cubic lattice with spacing $x = \sqrt[3]{V}$, with each atom trapped in a roughly parabolic potential well with curvature $u''$. The Helmholtz free energy function will take the form $A(T,V) = -3T\log T + \frac{3}{2}T\log u''(x) + 3u(x)$. Hence,

$$\left(\frac{dS}{dT}\right)_V = -\left(\frac{d^2A}{dT^2}\right) = \frac{3}{T}; \quad \left(\frac{dP}{dT}\right)_V = -\left(\frac{d^2A}{dVdT}\right) = \frac{u'''}{2x^2u''}; \quad \left(\frac{VdP}{TdS}\right)_V = \frac{xu'''}{6u''}$$

We can use these data to estimate the Dissipation Number for the earth:

$$\int dR\, \frac{\alpha g}{C_P/m} = 3.0\int dR\, \frac{g}{u^2} \approx 0.65$$

The last key parameter is the electrical conductivity, which is particularly elusive but can be related to measured thermal conductivity using the Wiedemann-Franz Law. The conductivities should scale under compression as $\rho^{-1/3}$, based on the electron density, mean free path, and Fermi velocity.

The thermal conductivity of molten iron has not been published, so we will take the value for solid iron just below the melting point as an upper bound. (One might expect the conductivities to drop sharply at the melting point because of increasing disorder. Melting can be viewed as a catastrophic softening of transverse phonon modes, and the electrons' mean free path is determined by electron-phonon scattering. Electrons couple primarily to longitudinal phonons when the electron-ion interaction is a central potential, but iron's conduction electrons reside in D-orbitals, and they would also couple to transverse phonons.)

Common estimates of electric conductivity in the core vary from $10^5$ to $10^6$ S/m, where S = siemens = mho. Our upper-bound, 0.6 MS/m, predicts a free decay time of 30,000 years for the dipole field in the absence of a dynamo effect. Taking ~ 4 gauss for the external dipole field extrapolated into the core, we find $\beta \equiv \frac{\sigma}{\rho} B_{ext}^2 \sim$ 1/day, which is tantalizingly close to the rotation period.

*Ab initio* calculations (using density functional theory to estimate the electronic energy and Monte-Carlo methods to compute the partition function) have not yet attained impressive accuracy or credibility.

**Conditions in the Sun**

The sun is composed of fully ionized hydrogen and helium, virtually ideal gases. Within the convective (outer) zone, the relationship is adiabatic: $\rho \sim P^{3/5}$ and $T \sim P^{2/5}$. The speed of sound is $u^2 = (dP/d\rho)_S = \frac{5}{3} P/\rho$, and the Grüneisen parameter is 2/3.

**Conditions in the Sun's Convective Zone**

| Parameter | Symbol | Outer Boundary | Inner Boundary | Units |
|---|---|---|---|---|
| Radius | R | 696,000 | ~ 500,000 | km |
| Gravity | g | 274 | 545 | m/s$^2$ |
| Pressure | P | ~ 0 | 6.5 E12 | Pa |
| Density | $\rho$ | ~ 0 | 0.21 | g/cc |
| Temperature | T | 5800 | 2.3 E6 | Kelvin |
| Luminosity | -- | 3.86 E26 | 3.86 E26 | W |
| Electrical conductivity | $\sigma$ | -- | ~ 2 E7 | S/m |

We can use these relationships to calculate the Dissipation Number for the sun:

$$\tfrac{2}{3}\int dR\, g/u^2 = \tfrac{2}{5}\int dP/P = \Delta(\log T) \approx 6$$

The electrical conductivity scales as $\sigma = ne^2\tau/m_e$, where $n$ is the number density of electrons, and $\tau$ the relaxation time. The standard formula for energy loss due to Coulomb scattering (or just dimensional analysis) suggests that $\tau \sim T^{3/2} m_e^{1/2} / Z^2 e^4 n_Z$, with logarithmic corrections, and it follows that $\sigma \sim T^{3/2} \sim P^{3/5} \sim \rho$ along the adiabat. The Spitzer-Härm formula (based on the Boltzmann transport equation) for hydrogen is

$$\sigma \approx \frac{(0.0153\,\text{mho/m})}{\log(\Lambda)} \left(\frac{T}{\text{Kelvin}}\right)^{3/2} \quad \text{where} \quad \Lambda^2 \sim \frac{(kT)^3}{4\pi e^6 n_{e+p}}$$

(Spitzer & Härm did not give a precise expression for $\Lambda$, but it is to be interpreted as follows: $\Lambda$ is the ratio of the Debye screening length to the impact parameter that produces a 90-degree deflection, $b_o \approx e^2/mv^2$.)

Taking 1-2 gauss for the external dipole field of the sun, measured at the poles, we find that $\beta/\Omega$ is roughly $10^2$, vastly greater than for the earth. (The more commonly cited value of 100 gauss refers to the field strength in solar prominences, and this distinction has been a source of confusion.)